\documentstyle[preprint,aps,eqsecnum]{revtex}

\begin{document}

\title{\bf The Role of Fermions in Bubble Nucleation}

\author{\bf D.G. Barci$^{a)}$}

\address{Instituto de F\'\i sica \\ Universidade Estadual
do Rio de Janeiro \\ Rua S\~ao Francisco Xavier, 524 \\
Maracan\~a, Rio de Janeiro, RJ, 20559-900, Brasil}

\author{\bf E.S. Fraga$^{b)}$ and C.A.A. de Carvalho$^{c)}$}

\address{Instituto de F\'\i sica \\ Universidade Federal
do Rio de Janeiro \\ C.P. 68528, Rio de Janeiro, RJ, 21945-970, Brasil}

\date{}

\maketitle

\begin{abstract}

We present a study of the role of fermions in the decay
of metastable states of a scalar field via bubble
nucleation. We analyze both one and three-dimensional systems by
using a gradient expansion for the calculation of the fermionic determinant.
The results of the one-dimensional case are compared to the exact results
of previous work.

\end{abstract}

\section{Introduction}

The study of the mechanism of decay of metastable systems via nucleation 
of thermally activated bubbles finds a wide range of applicability in Condensed 
Matter physics. In fact, the formalism developed in \cite{CAC1} and \cite{FC} 
fits naturally in
the study of the dynamics of defects in one-dimensional conducting polymers 
\cite{yulu,BCF}. Applications to three-dimensional systems are also numerous, 
from the behaviour of binary liquid mixtures \cite{langer,Notes} to problems 
in Optics, such as atomic ionization in ultra-strong laser fields \cite{Davi},
and even Baryogenesis \cite{baryon}.

In a previous article \cite{FC}, we
studied a one-dimensional system of interacting fermions and bosons that
started in a metastable vacuum and gradually decayed to the true one. We 
investigated the nucleation of bubbles of true vacuum inside
the false one via thermal activation. We analyzed the stability of these 
bubbles and
calculated the decay rate as a function of time, in the presence of a
finite density of fermions, at finite temperature. The inclusion of
fermions proved to have remarkable effects as it changed qualitative features
of the physics of metastability \cite{FC}.

The fact that we were working with a one-dimensional system allowed us
to solve the problem exactly by means of inverse scattering methods.
However, as we move to higher dimensions, this mathematical tool is
no longer available and we are compelled to make
approximations to evaluate the fermionic 
determinant that encodes the fermionic effects on the system. This may be
accomplished by the use of a functional gradient expansion for the
determinant. Throughout this work, we perform our calculations for both
one and three-dimensional systems, so that we can compare the results
of our approximations on the former with the exact results of \cite{FC}.
In fact, our approximation is able to detect the phenomenon of ``quantum
stabilization'' for both one and three-spatial dimensions.

The paper is organized as follows: in section II, we present the model 
considered; in section III, we perform the gradient expansion for the
fermionic determinant; in section IV, we obtain the form of the sphaleron
solution that includes fermionic effects;
in section V, we include the presence of the gap and two bound states
of fermions; in section VI, we study the stability of bubble-like
solutions; in section VII, we comment on the methods
and present our conclusions.

\section{The model}

The model Lagrangian has the following form

\begin{equation}
\label{lagrangeana2}
{\cal L}={1\over 2}(\partial_\mu\phi)(\partial^\mu\phi)-[V(\phi) -
V(\phi_2)] + \bar\psi_a(i\gamma^{\mu}\partial_{\mu}-\mu-g\phi)\ \psi_a
\end{equation}

\noindent
where $\mu$ is the bare mass of the fermions, $g$ is the coupling
constant, $\psi_a(x)$ is the fermion field, $a$ denotes fermion
species, $\phi(x)$ is a scalar field and $\phi_2$ is a local minimum of
the potential 

\begin{equation}
\label{potencial2}
V(\phi)={g^2\over 2}(\phi-\phi_0)^2\left(\phi+\phi_0+{2\mu\over
g}\right)^2 + j\phi 
\end{equation}

\noindent
where $\phi_0$ is a constant and $j$ is an external current, 
responsible for the asymmetry of the potential even in the 
purely bosonic case. (We may find a physical realization of this form 
of potential in the description of conducting polymers \cite{yulu,BCF};
see Fig. 1).

We will be interested in the effects of fermions on
the bosonic field. Thus, in order to construct an effective theory for
bosons, we must integrate over fermions to obtain an effective action.
However, integrating over fermions implies calculating the following
determinant

\begin{equation}
\label{det}
S_F=-ln[det(i\gamma^{\mu}\partial_{\mu}-\mu-g\phi)]=-tr[ln(i\gamma^{\mu}
\partial_{\mu}-\mu-g\phi)] 
\end{equation}

\noindent
In one spatial dimension, it is possible to perform an exact calculation of
(\ref{det}) by making use of inverse scattering methods \cite{FC,CAC3,Novikov}.
For three spatial dimensions, we must resort to approximations.

\section{Gradient expansion for the fermion determinant}

After the fermionic integration, we may rewrite the effective action as

\begin{equation}
S_{eff}[\phi]=\int d^{\nu}x\left\{{1\over 2}(\partial_\mu\phi)(\partial^\mu\phi)-
[V(\phi) - V(\phi_2)]\right\} - tr[ln(i\gamma^{\mu}\partial_{\mu}-\mu-g\phi)]
\end{equation}

\noindent
where $\int d^{\nu}x \equiv \int^T_0 dt \int d^D x$. The field configuration 
that extremizes (3.1) must satisfy the Euler-Lagrange equation

\begin{equation}
\Box\phi=-\frac{\partial V}{\partial\phi}-
Sp<x|\frac{1}{i\gamma^{\mu}\partial_{\mu}-\mu-g\phi}|x>
\end{equation}

\noindent
where $Sp$ means trace over the spin degrees of freedom.

We may calculate the Green function that appears in (3.2) by using a functional 
gradient expansion \cite{Ait}. This means that we will focus on the long distance 
(small momentum) properties of the theory. To do so, we use the identity  
\cite{sakita,barci}(see Appendix A)

\begin{eqnarray}
{\cal G}(x,x) &\equiv& 
Sp<x|\frac{1}{i\gamma^{\mu}\partial_{\mu}+M(x)}|x>=\nonumber \\
&=& Sp \int\frac{d^{\nu}p}{(2\pi)
^{\nu}} \frac{1}{\gamma^{\mu}p_{\mu}+M(x)} \sum_{n=0}^{\infty} (-1)^n\left( \Delta M
(\frac{1}{i}\frac{\partial}{\partial p},x) \frac{1}{\gamma^{\mu}p_{\mu}+M(x)}
\right)^n
\label{apA}
\end{eqnarray}

\noindent
where

\begin{equation}
\Delta M
(\frac{1}{i}\frac{\partial}{\partial p},x)=\partial_{\mu}M(x)\frac{1}{i}
\frac{\partial}{\partial p_{\mu}} + \frac{1}{2} \partial_{\mu}\partial_{\nu}
M(x)(\frac{1}{i})^2 \frac{\partial^2}{\partial p_{\mu} \partial p_{\nu}} + ...
\end{equation}

Keeping terms up to second order in the derivatives and explicitly
performing the integrals, we obtain

\begin{equation}
\label{fg}
{\cal G}(x,x) = \alpha_{\nu}M^{\nu-1}+
\beta_{\nu}M^{\nu-4} \Box M
\end{equation}

\noindent
where $\alpha_{\nu}$ and $\beta_{\nu}$ are functions of the number of space-time
dimensions:

\begin{equation}
\alpha_{\nu} \equiv \frac{\pi^{-\nu/2}}{2^{\nu+1}} \Gamma(1-\nu/2)\nonumber
\end{equation}

\begin{equation}
\beta_{\nu} \equiv \frac{\pi^{-\nu/2}}{2^{\nu-1}\Gamma(\nu/2)} \left[
\frac{\Gamma(\nu/2)}{4!}\Gamma(4-\nu/2) - \frac{\Gamma(1+\nu/2)}{2!2!}
\Gamma(3-\nu/2) + \frac{\Gamma(2+\nu/2)}{4!}\Gamma(2-\nu/2) \right]
\end{equation}

For $\nu = 2$ and $\nu = 4$, these functions are divergent. 
Nevertheless, we may absorb these divergences by a suitable redefinition of the
free parameters of the model. After a redefinition of $g$, $\mu$ and $j$ in (2.2)
and remembering that, in our case, $M(x)=-(\mu + g\phi)$, we may explicitly 
calculate the equation of motion (3.2).

For $\nu = 2$, equation (3.2) reads (up to order $g^2$)

\begin{equation}
\Box\phi = -\frac{\partial V}{\partial\phi} \left[ \frac{(\mu+g\phi)^2}
{(\mu+g\phi)^2 + \frac{g^2}{6\pi}} \right]
\end{equation}

\noindent
and, for $\nu = 4$, 

\begin{equation}
\Box\phi = -\frac{\partial V}{\partial\phi}
\end{equation}

The part of the effective action, $S_F$, associated with the fermionic
determinant has the following form \cite{nota}

\begin{equation}
S_F = -g \int d^{\nu}x \int[D\phi]~ {\cal G}(x,x)
\end{equation}

\noindent
with the matrix element given by (3.5). Thus, for $\nu=2$,

\begin{equation}
S_F= -\frac{g}{12\pi} \int d^2x \frac{\Box\phi}{(\mu + g\phi)}
\end{equation}

We note that, in the case of $\nu = 4$, the only effect of the fermions
is (to this order of approximation) to renormalize the free parameters of the 
theory. We will see, in section V, that this situation is completely 
changed if we include a gap and bound states in the fermionic spectrum 
(see Fig. 4).

\section{The sphaleron solution}

Based on the results of our previous work, we shall look for 
{\it sphaleron}-like solutions to the equations of motion obtained in
the last section. A {\it sphaleron}-like solution corresponds to a field
configuration that starts at the false vacuum $\phi_2$, almost reaches
the true vacuum $\phi_1$, and returns to $\phi_2$, i.e., it looks like
a droplet \cite{CAC1,FC,BCF,CAC2}.

\subsection{The $\nu=2$ case}

In this case, the equation of motion is (3.8). Defining $\varphi
\equiv \phi+\mu/g$, using the thin-wall approximation \cite{CAC1} 
(i.e., considering a droplet whose radius is much larger than its wall 
thickness) and imposing the boundary conditions

\begin{equation}
\label{profile}
\varphi_{sph}(x\rightarrow \pm \infty)\rightarrow \varphi_2
\end{equation}

\begin{equation}
\frac{d\varphi_{sph}}{dx} (x\rightarrow \pm \infty)\rightarrow 0
\end{equation}

\noindent
we obtain

\begin{equation}
\label{sph1}
\varphi_{sph}=\varphi_2-\phi_P [tanh(\xi+\xi_0)-tanh(\xi-\xi_0)]
\end{equation}

\noindent
where

\begin{equation}
\phi_P \equiv \left[ \frac{(3\varphi^2_2-\varphi^2_0)}{2} \right]^{1/2}
\end{equation}

\begin{equation}
\xi \equiv g\phi_P(x-x_{c.m.})
\end{equation}

\begin{equation}
\xi_0 \equiv \frac{1}{2}cosh^{-1}\left(\sqrt{\frac{2}{\frac{\varphi_0^2}{
\varphi_2^2}-1}} \right)
\end{equation}

\noindent
and where we have assumed $\mu/g >>1$ (neglecting corrections of order
$O((g/\mu)^2)$), consistent with the conditions of 
validity of the gradient expansion, in order to obtain a closed form
for the function that represents the sphaleron.

The parameter $x_{c.m.}$ reflects the translational invariance of the
equation of motion and $\xi_0$ is related to what is usually called
the radius of the sphaleron (see Fig. 2), being extremely important
in the analysis of stability (see section VI).

\subsection{The $\nu=4$ case}

In this case, the equation of motion is (3.9). Thus, assuming a solution with 
radial symmetry, we have, in the thin-wall approximation,

\begin{equation}
\label{sph3}
\varphi_{sph}(\tilde\xi)=\varphi_2-\tilde\phi_P[tanh(\tilde\xi+\tilde\xi_0)-
tanh(\tilde\xi-\tilde\xi_0)],~~~~~~\tilde\xi \geq 0
\end{equation}

\noindent
where we have imposed the following conditions

\begin{equation}
\tilde\phi_P \equiv \left( \frac{3\varphi_2^2-\varphi_0^2}{2}\right) ^{1/2}
\end{equation}

\begin{equation}
\tilde\xi \equiv g\tilde\phi_Pr
\end{equation}

\begin{equation}
\tilde\xi_0 \equiv \frac{1}{2}cosh^{-1} \left( \frac{2\varphi_2}{\sqrt{
2(\varphi_0^2-\varphi_2^2)}} \right)
\end{equation}

\noindent
and

\begin{equation}
\label{profile3}
\varphi_{sph}(r\rightarrow \infty)\rightarrow \varphi_2
\end{equation}

\begin{equation}
\frac{d\varphi_{sph}}{dr} (r\rightarrow \infty)\rightarrow 0
\end{equation}

Therefore, we have a true vacuum bubble, of radius $\tilde\xi_0$, centered
at the origin (see Fig. 3). The situation is the same as we would have 
encountered in a purely bosonic system \cite{CAC1}.

\section{Inclusion of a gap and bound states}

In the previous section, we have shown that a bubble-like configuration
still satisfies the modified equations of motion (In fact, in the $\nu=4$ case,
we have no changes at all). These bubbles play the role of a background field
in the determination of the fermionic spectrum. In fact, in some cases, 
instead of a simple
continuum, we may have the presence of a gap and pairs of bound states \cite{FC,CAC3} 
(we will consider, for simplicity, just one pair). Therefore, if we have a richer
spectrum (see Fig. 4), the calculation of the fermionic determinant, performed in
section III, should be reviewed to deal explicitly with the bound states. The 
calculation that led us to (3.11) is the
result of a gradient expansion approximation for the total trace (2.3), where we 
sum over all continuum and bound states. However,
as we are interested in an effect that depends crucially on the relative 
occupation of the bound states \cite{FC}, we should consider a finite density
of fermions, i.e., a partial trace. Thus, instead of summing over all fermionic
momenta (the restriction of small momenta is over the bosonic fields only), we
shall sum them only up to the (occupied) top bound state (see Fig. 4(c)). 

Therefore, the complete effective action has the following form

\begin{eqnarray}
S_{eff}[\phi]&=&\int d^{\nu}x ~ \left\{\frac{1}{2}(\partial_\mu\phi)(\partial^\mu\phi)-
[V(\phi)-V(\phi_2)]\right\}-\nonumber \\
&-& \frac{g}{2} \int d^{\nu}x \int [D\phi]~ {\cal G}(x,x) -\nonumber \\
&-& n_+ g \int_0^T dt_1 \int d{\vec x} \int_0^T dt_2 \int d{\vec y} \int [D\phi] ~ \psi_B^*({\vec x},t_1,[\phi_{sph}]) \times\nonumber \\
&\times& {\cal G}(x,y) \psi_B({\vec y},t_2,[\phi_{sph}])
~~~~~~~~~~~
\end{eqnarray}

\noindent
where the first term is the bosonic contribution, the second term 
represents the ``Dirac sea'' (including antiparticle bound states) 
and the last one is due to the particle bound states.
$\psi_B$ is the wavefunction of a bound state and $n_{+}$ is its 
occupation number (``doping'').

The only term that remains to be calculated  is the last one. Assuming that 
the occupation of the bound states will not affect in an appreciable way the 
form of the bubble (in the one-dimensional case, this is an exact result 
\cite{FC,CAC3}), we may rewrite this term as (see Appendix B)

\begin{equation}
S_{bound}=n_+ gT \int d{\vec x} \rho({\vec x}) \phi_{sph}({\vec x})
\label{apB}
\end{equation}

\noindent
where $\rho({\vec x})=\psi^*_B({\vec x}) \psi_B({\vec x})$ is the normalized density
of probability distribution of the bound charge. However, it is already known
\cite{yulu,CAC3} that the charge associated with a bubble tends to concentrate
on its surface in a gaussian-like way. For our purposes, we will assume that
a delta-like distribution will be a reasonable approximation.

\subsection{The $\nu=2$ case}

In this case, we may write the density $\rho$ as

\begin{equation}
\rho=\frac{1}{2} [\delta(\xi-\xi_0)+\delta(\xi+\xi_0)]
\end{equation} 

The energy of the bubble, $E=-S_{eff}/T$, as a function of the radius $\xi_0$, has 
the following form:

\begin{eqnarray}
E(\xi_0)&\approx& \int_{-\infty}^{+\infty} \frac{d\xi}{g\phi_P} 
\left\{ \frac{1}{2} 
\left[ 1+ \frac{1}{12\pi} \left( \frac{g}{\mu} \right)^2 \right]
(g\phi_P)^2 \left( \frac{d\phi_{sph}}{d\xi} \right)^2 + 
[V(\phi_{sph})-V(\phi_2)]\right\} -\nonumber \\
&-& n_+ tanh(2\xi_0)
\end{eqnarray}

\noindent
where we have incorporated the contribution of the Dirac sea in the first term, by 
making use of the equation of motion. In fact, the contribution of the continuum is
of order $O((g/\mu)^2)$ and may be neglected to this order of approximation. The
only relevant contribution of fermions comes from the bound states.

\subsection{The $\nu=4$ case}

In this case, we may write the density $\rho$ as

\begin{equation}
\rho=\frac{(g\tilde\phi_P)^2}{4\pi\tilde\xi_0^2} \delta(\tilde\xi-\tilde\xi_0)
\end{equation}

The energy of the bubble as a function of the radius $\tilde\xi_0$ has the following 
form:

\begin{eqnarray}
E(\tilde\xi_0)&\approx& \int_0^{\infty} 4\pi \frac{\tilde\xi^2}{(g\tilde\phi_P)^2} 
\frac{d\tilde\xi}{(g\tilde\phi_P)} ~ \left\{{1\over 2}(g\tilde\phi_P)^2 \left( 
\frac{d\phi_{sph}}{d\tilde\xi} \right)^2 + 
[V(\phi_{sph})-V(\phi_2)]\right\}-\nonumber \\
&-& n_+ tanh(2\tilde\xi_0)
\end{eqnarray}

\section{Stability of the bubbles}

To analyze the stability of the bubble-like solutions obtained in section IV,
we shall study the behaviour of the the energy of the
bubble as a function of the bubble radius $\xi_0$, now considered as a 
dynamical variable, $s$. The results, for the cases $\nu=2$ and $\nu=4$ 
obtained in last section, are plotted in Fig. 5 and Fig. 6, respectively.
Our results for the $\nu=2$ case should be compared with the exact results,
previously obtained by using inverse scattering methods \cite{FC} (see Figs. 
5(b) and 5(c)). In this way, we may control our approximations. 
In fact, the observation of Fig.5 shows that our approximation preserves the ``quantum stabilization'' brought about by fermions. Nevertheless, 
we find a quantitative difference between the approximate and the exact results 
due, mainly, to our naive delta-like approximation for the fermionic density. In 
order to improve on this approximation, one may use a gaussian-like pattern or
even the exact fermionic density in the presence of the sphaleron. However, 
in doing so, one can no longer use the simple analytic form of the previous section.

\section{Conclusions and comments}

The aim of this work was to understand the role played by fermions in the decay
of metastable states of a scalar field via bubble nucleation. Our basic approximation
was a gradient expansion for the fermionic determinant that proved to work well
in a comparison with exact results previously obtained for one-dimensional
system. From our results, it is clear that the effects of fermions in an arbitrary
number of space dimensions are due, almost
exclusively, to the relative occupation of bound states. The states of the continuum 
contribute only to order $O((g/\mu)^2)$, and may be neglected.(This is in agreement
with results obtained for an analogous problem in QCD \cite{QCD}.).
For $\nu=2$ and $\nu=4$, 
the new charged bubbles have the same functional form as the purely bosonic ones, except for a reparametrization, 
and have their stability drastically modified by the fermions: besides unstable
bubbles, we find metastable bubbles as a result of a ``quantum stabilization'' brought
about by the fermions. For $\nu=2$, this result was obtained exactly 
in Ref.\cite{FC}. This striking feature may, in principle, be measured in some 
realistic systems. The two-dimensional case finds a natural application in the 
physics of linearly conducting polymers \cite{yulu,BCF,CAC3}. The results for the four-dimensional 
case may have important consequences for the physics of Baryogenesis and in some 
problems in Optics  \cite{Davi,baryon}

From the results for the energy of the bubbles as a function of their radii, it is
possible to calculate decay rates for the metastable states of the scalar field as
explicit functions of time. This may be implemented by using the formalism presented
in \cite{CAC1} and \cite{FC} and allows for a non-equilibrium description of the decay 
process for both $\nu=2$ and $\nu=4$ cases.

\bigskip\bigskip

{\bf Acknowledgements}

The authors acknowledge CNPq and FUJB/UFRJ for finantial support. C.A.A.C. would
like to thank ICTP for kind hospitality during part of this work.

\bigskip\bigskip

\appendix 
\section{Gradient Expansion of the Fermion Density}

The complete derivation of identity (\ref{apA}) can be found in
 ref. \cite{sakita}. For  completeness,
 we present here the main steps of its demonstration.  

The density of fermions at a given point $x_0$, in a background field $M(x)$, has
the following form
 
\begin{equation}
 \rho(x_0)\equiv{\cal G}(x_0,x_0)\equiv Sp<x_0|\frac{1}{i \gamma^{\mu}
\partial_{\mu} +M(\hat{x})}|x_0>
\end{equation}

\noindent
where $|x_0>$ is a position eingenstate with eigenvalue $x_0$, and $Sp$ is the trace over spin degrees of freedom. 

In momentum representation, this expression takes the form
 
\begin{equation}
\rho(x_0)= Sp\int \frac{d^\nu p}{(2\pi)^\nu}e^{ix_0^\mu p_\mu}\frac{1}{\gamma^{\mu}
p_{\mu} +M(\hat{x})}e^{-ix_0^\mu p_\mu}
\end{equation}
In order to transfer the $x_0$ dependence to $M(\hat{x})$, we make a unitary transformation  that eliminates the exponentials from  the last expression:
\begin{equation}
\rho(x_0)= Sp\int \frac{d^\nu p}{(2\pi)^\nu}\frac{1}{\gamma^{\mu}p_{\mu}
 +M(\hat{x}+x_0)}
\label{gg}
\end{equation}  

Note that $x_0$ in $M(\hat{x}+x_0)$ is a  {\em c-number,
not an operator}.

We can now expand $M(\hat{x}+x_0)$ around $x_0$:

\begin{equation}
M(\hat{x}+x_0)=M(x_0)+\Delta M(x_0,\hat{x})
\label{deltam}
\end{equation}

\noindent
where

\begin{equation}
\Delta M(x_0,\hat{x})=\partial_\mu M(x_0) \hat{x}^\mu+
\frac{1}{2}\partial_\mu\partial_\nu M(x_0) \hat{x}^\mu\hat{x}^\nu+\ldots
\end{equation}

Using (\ref{deltam}) in (\ref{gg}) and considering that, in momentum space, $\hat{x}_\mu=\frac{1}{i}\frac{\partial~}{\partial p_\mu}$, we have

\begin{equation}
\rho(x_0)= Sp\int \frac{d^\nu p}{(2\pi)^\nu}\frac{1}{\gamma^{\mu}
p_{\mu} +M(x_0)+\Delta M(x_0,\frac{1}{i}\frac{\partial~}{\partial p_\mu})}
\label{gg2}
\end{equation} 

We can factor out $\frac{1}{\gamma^{\mu}p_{\mu}+M(x_0)}$ obtaining: 

\begin{equation}
\rho(x_0)= Sp\int \frac{d^\nu p}{(2\pi)^\nu}
\frac{1}{\gamma^{\mu}p_{\mu} +M(x_0)} \left[
1+\Delta M(x_0,\frac{1}{i}\frac{\partial~}{\partial p_\mu})\times
\frac{1}{\gamma^{\mu}p_{\mu} +M(x_0)} \right]^{-1} 
\label{gg3}
\end{equation}

The idea is to expand the last parenthesis in powers of
$\Delta M/(\gamma^{\mu}p_{\mu}+M)$. This means that we consider
 smooth backgrounds $M(x_0)$, so 
that $|\partial_\mu M(x_0)|<|M(x_0)|$. In momentum space, this condition 
implies $|k_\mu \tilde{M}(k)|<|\tilde{M}(k)|$, where $k<1$ is the momentum 
transferred  from the fermions to the boson background. With this expansion 
we finally obtain:

 \begin{equation}
\rho(x_0)= Sp\int \frac{d^\nu p}{(2\pi)^\nu}
\frac{1}{\gamma^{\mu}p_{\mu} +M(x_0)}
\sum_{n=0}^{\infty} (-1)^n
\left(\Delta M(x_0,\frac{1}{i}\frac{\partial~}{\partial p_\mu})\times
\frac{1}{\gamma^{\mu}p_{\mu} +M(x_0)}\right)^n 
\label{gg4}
\end{equation}

\section{Energy of bound states}

In this appendix we derive equation (\ref{apB}). 

From (5.1) we have 

\begin{eqnarray}
S_{bound}&=&-n_+ g \int dt_1\int d\vec{x}\int dt_2
\int d\vec{y}\int[D\phi]\times\nonumber \\
&\times&\psi^{\dagger}_B(x,t_1,[\phi_{sph}]){\cal G}(x,y)\psi_B(y,t_2,[\phi_{sph}])
\label{a1}
\end{eqnarray}

\noindent
where $\psi_B$ is the bound-state wave function in the presence of a sphaleron, 
and can be written in the following form

\begin{equation}
\psi_B({\vec x},t,[\phi_{sph}])=\psi_B(\vec{x})e^{i E_B t}
\end{equation}

\noindent
where $E_B$ is the energy of the bound state. 

The Green function in (\ref{a1}) takes the form

\begin{equation}
{\cal G}(x,y)=\sum_{n}
\psi_n^{\dagger}(\vec{y})\psi_n(\vec{x}) e^{-i E_n (t_2-t_1)}
\label{gf}
\end{equation}
where $ E_n<\epsilon_F$ and
\begin{equation}
(i\gamma^{\mu}\partial_{\mu}+\mu+\phi)\psi_n(\vec{x})=E_n \psi_n(\vec{x})
\end{equation}

\noindent
The index $n$ may be discrete or continuous,  depending on
the  part of the spectrum we are considering. The sum must be done up to the 
Fermi energy $\epsilon_F$. 

Using (\ref{gf}) in (\ref{a1}), we obtain

\begin{eqnarray}
S_{bound}&=&-n_+ g \int d\vec{x}
\int d\vec{y}\int dt_1\int
 dt_2\int[D\phi]\sum_{n<\epsilon_F}\times\nonumber \\
&\times&\psi^{\dagger}_B(\vec{x})\psi_n(\vec{x})
\psi^{\dagger}_n(\vec{y})\psi_B(\vec{y}) ~ e^{-i(E_n-E_B)(t_2-t_1)}
\label{a2}
\end{eqnarray}

\noindent
The difficulty in evaluating this expression resides in the 
$\phi$-dependence of $\psi_n$. However, in the semiclassical context we 
 consider fields ``near'' the sphaleron configuration. We also 
assume that the occupation of the bound states will not affect essentially
the form of the bubble (this is an exact result in one spatial dimension 
\cite{FC}). Considering static fluctuations around the sphaleron, we can 
interchange the order of evaluation of the time and the  
$\phi$ integrals and, using the formula

\[
\int dt_1 \int dt_2 ~ e^{-i(E_n-E_B)(t_2-t_1)}=T\delta(E_n-E_B)
\]

\noindent
we obtain

\begin{equation}
S_{bound}=-n_+ g T\int d\vec{x}\int d\vec{y}\int[D\phi]~\psi^{\dagger}_B(\vec{x})\psi_B(\vec{x})
\psi^{\dagger}_B(\vec{y})\psi_B(\vec{y}) 
\label{a3}
\end{equation} 

\noindent
provided that $\epsilon_F>E_B$. 

Note that the $\delta(E_n-E_B)$ fixes the $\phi$-dependence of $\psi_n$, 
so that we can integrate in $\phi$ and then evaluate the expression in the 
sphaleron configuration. This procedure leads to 

\begin{equation}
S_{bound}=-n_+ g T\int d\vec{x}\int d\vec{y} ~\phi_{sph}(\vec{x}) \psi^{\dagger}_B(\vec{x})\psi_B(\vec{x})
\psi^{\dagger}_B(\vec{y})\psi_B(\vec{y}) 
\label{a4}
\end{equation} 

\noindent
In this form, the ${\vec x}$ and ${\vec y}$ integrations decouple and, using orthonormality of the wave functions, we finally arrive at 

\begin{eqnarray}
S_{bound}&=&-n_+ g T\int  d\vec{x}~ \phi_{sph}(\vec{x})\psi^{\dagger}_B(\vec{x})\psi_B(\vec{x})\nonumber \\
&=& -n_+ g T\int d\vec{x}~ \rho(\vec{x})\phi_{sph}(\vec{x})
\label{a5}
\end{eqnarray}

\noindent
which is precisely equation (\ref{apB}).

\bigskip\bigskip\bigskip

\underline{\bf Figure Captions:}

\vspace{3mm}

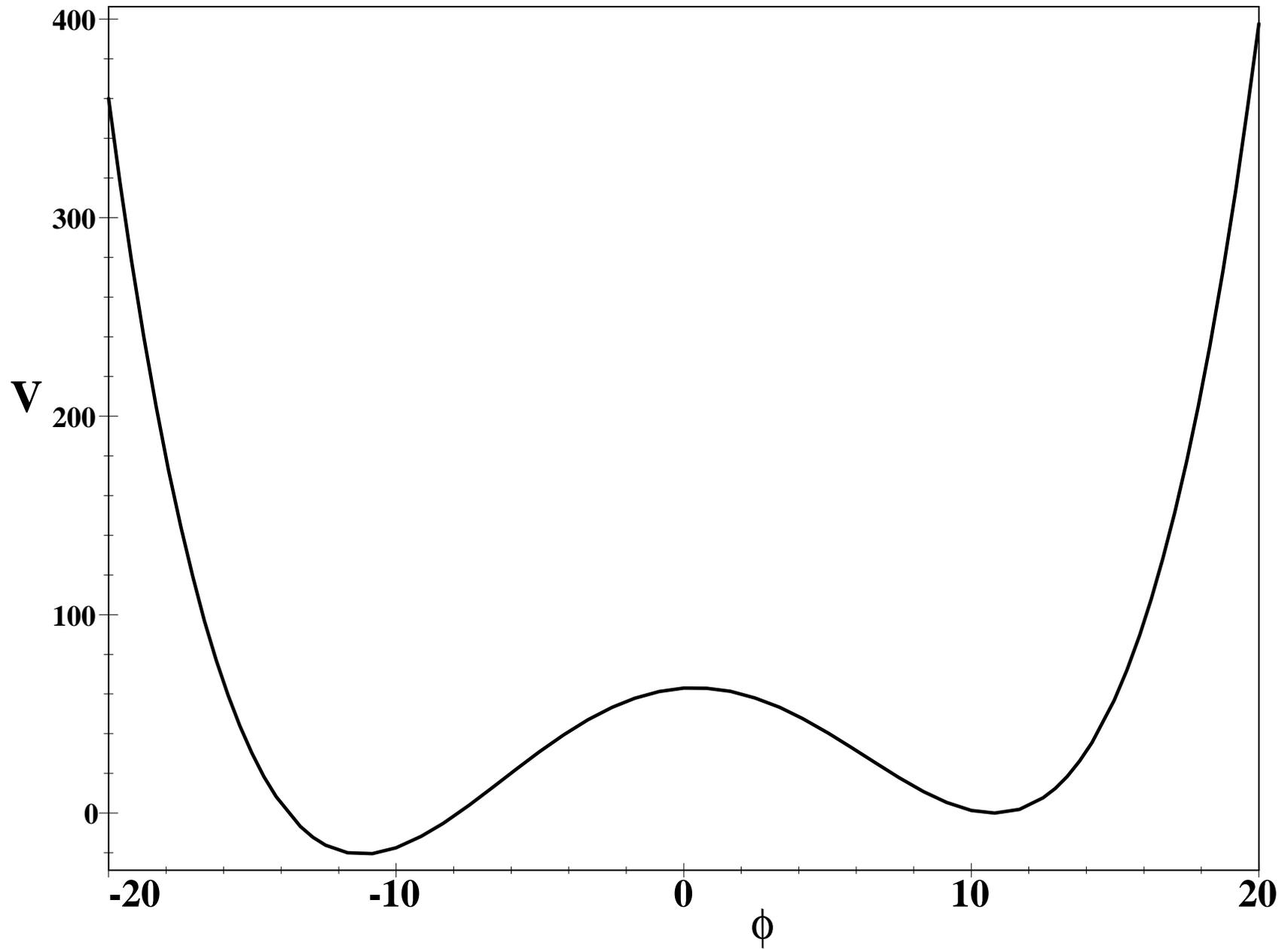
{\bf Figure 1}: Form of the potential $V(\phi)$.

\vspace{3mm}

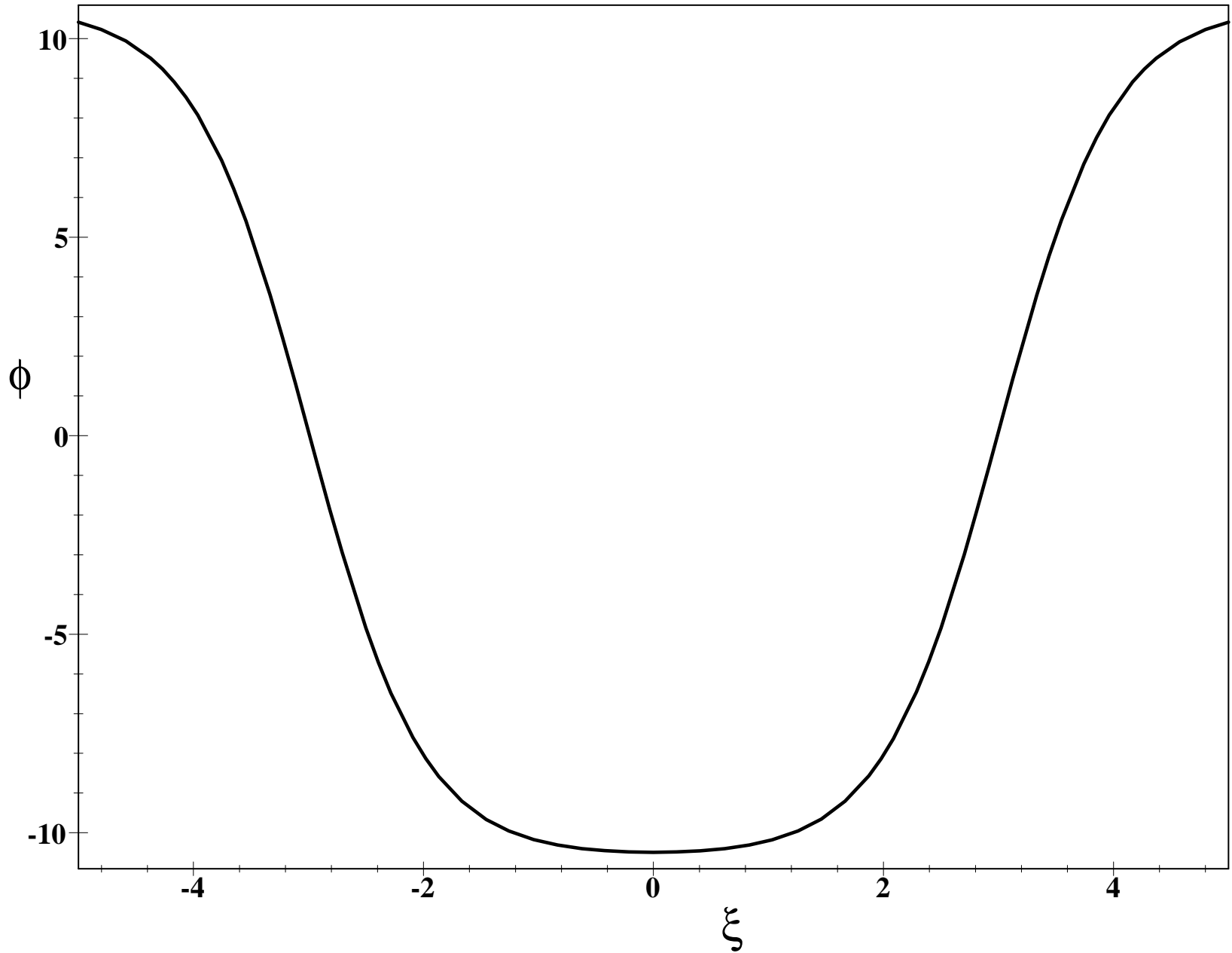
{\bf Figure 2}: The ``sphaleron'' solution for $\nu=2$.

\vspace{3mm}

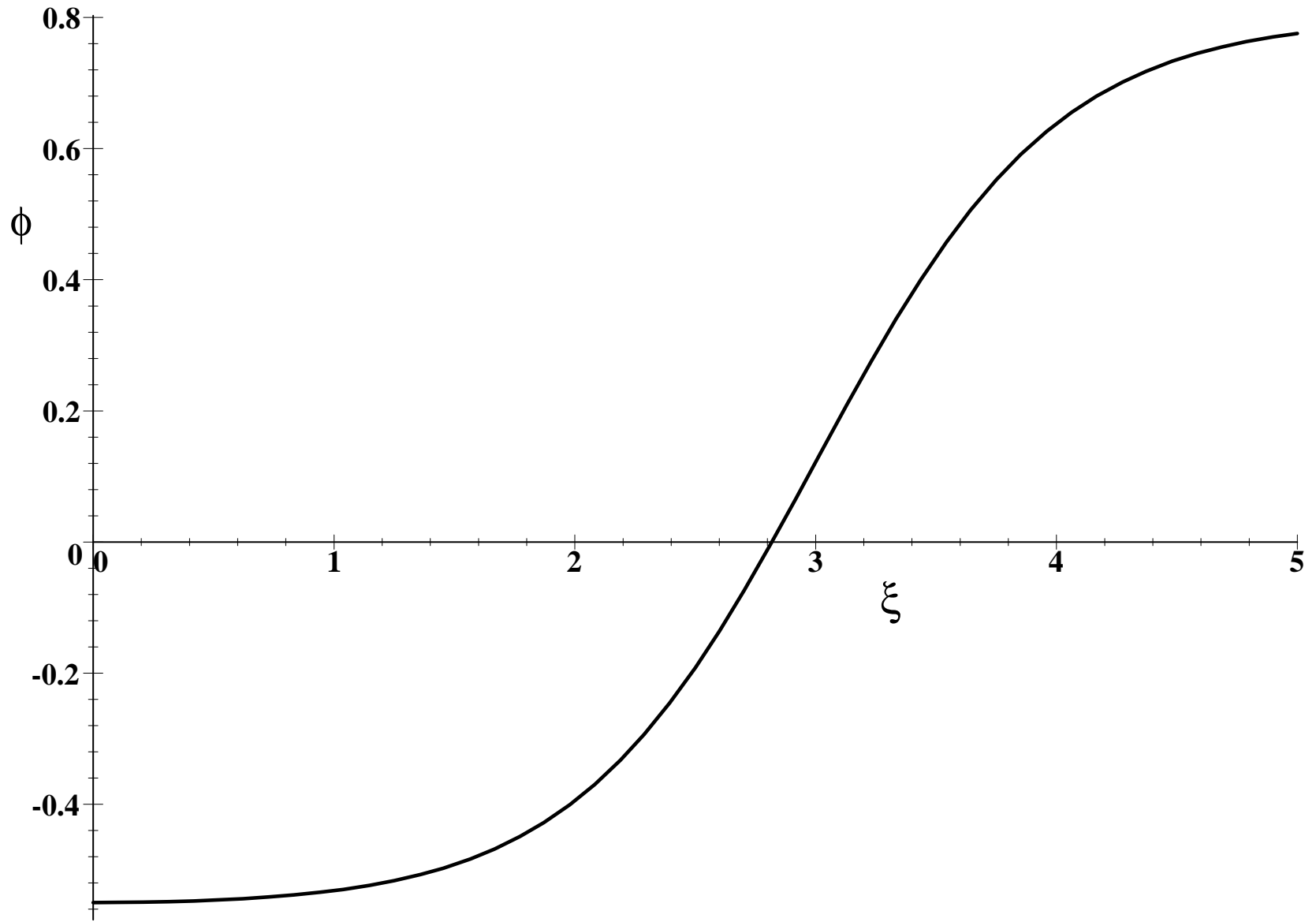
{\bf Figure 3}: The ``sphaleron'' solution for $\nu=4$.

\vspace{3mm}

{\bf Figure 4}: (a) Continuum (gapless)
 fermionic spectrum.
(b) Fermionic spectrum in the presence of a constant background. 
(c) Fermionic spectrum in the presence of a bubble-like background.
$\Omega$ is a cutoff (in polymer applications, it corresponds to the
bandwidth)

\vspace{3mm}

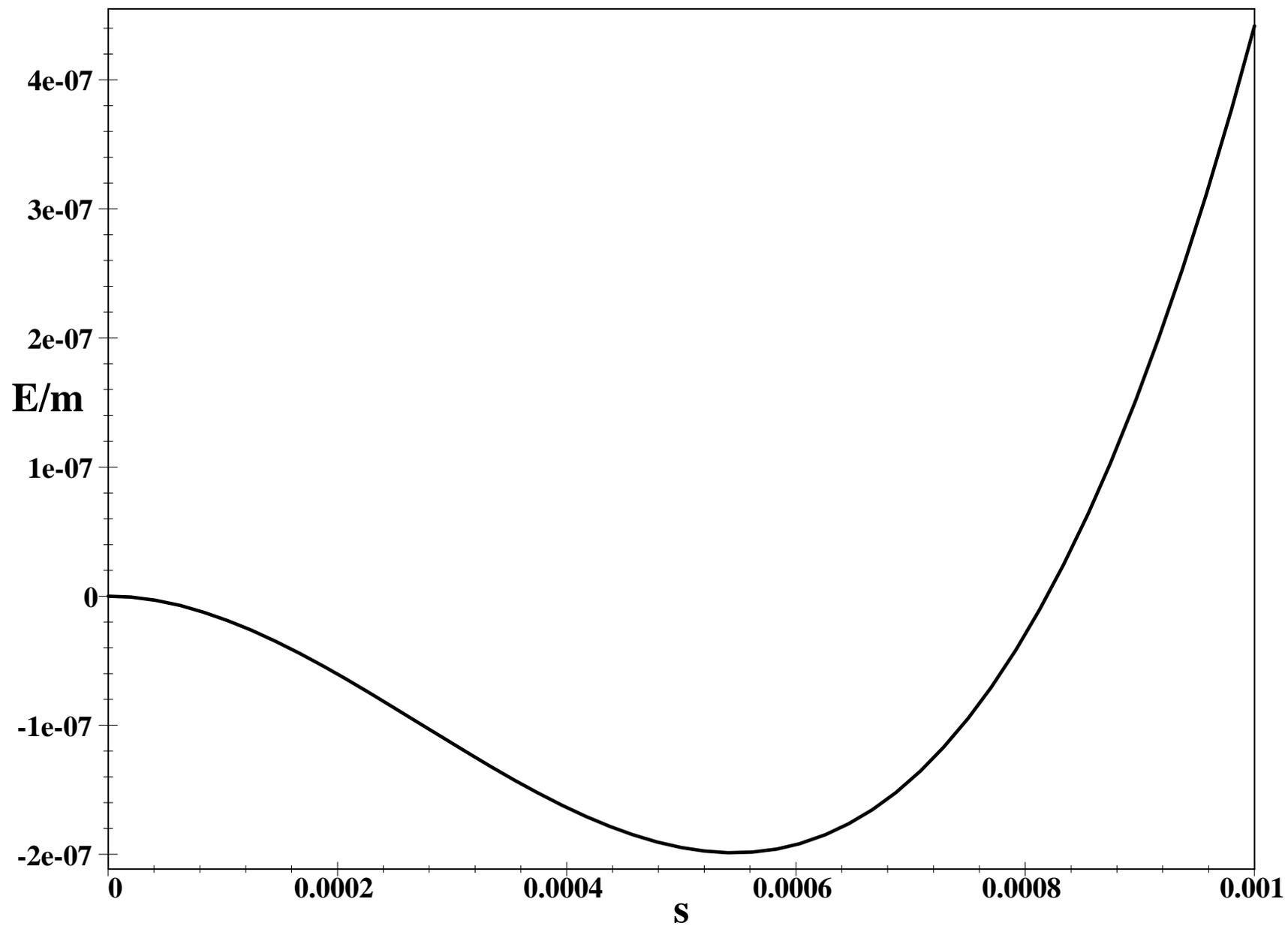
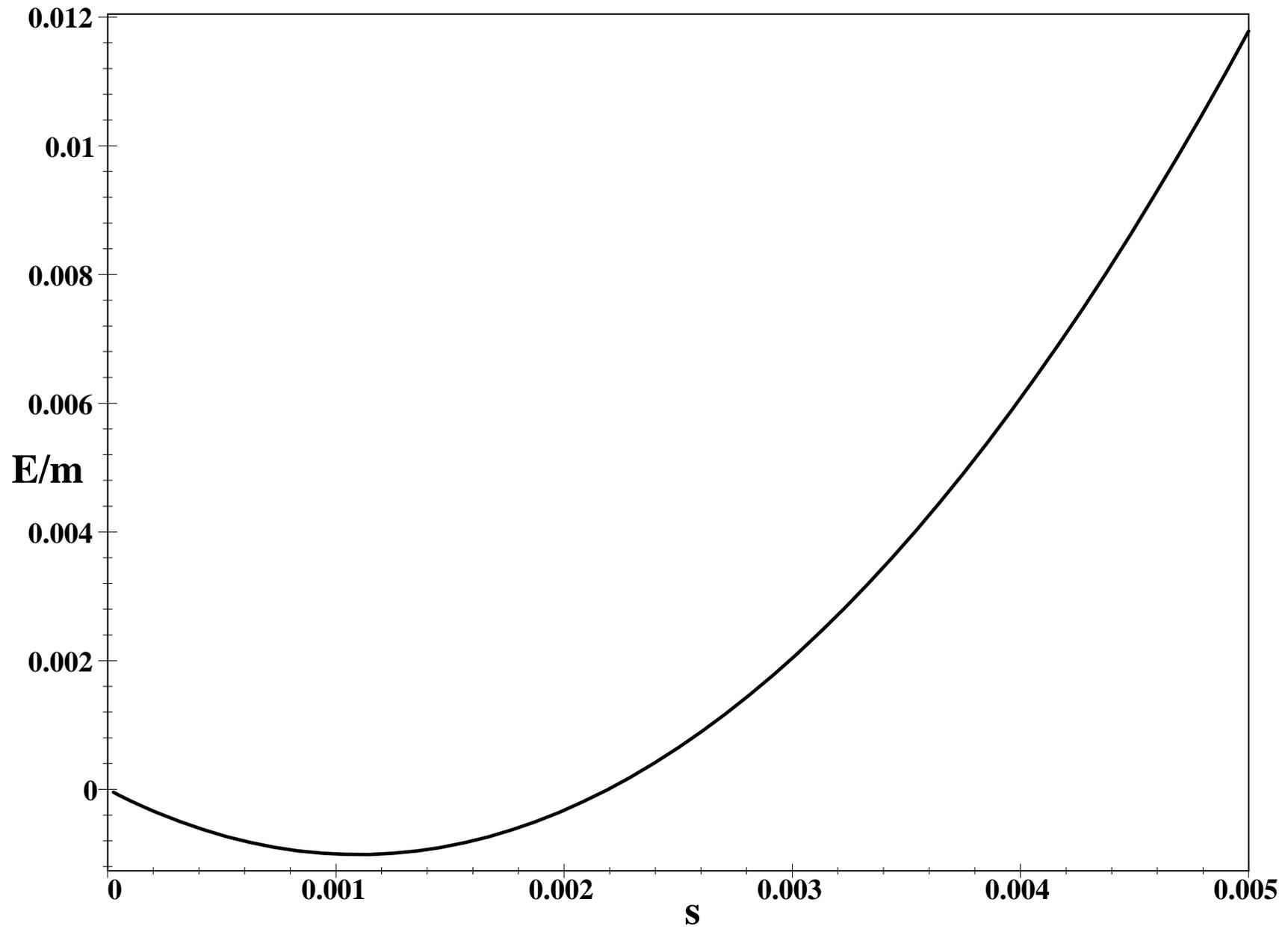
{\bf Figure 5}: Energy of the bubble as a function of the bubble radius
for $\nu=2$. (a) purely bosonic. (b) zoom of the interesting region for
the exact solution (including bound states). (c) zoom of the interesting 
region for the approximate solution (including bound states). 

\vspace{3mm}

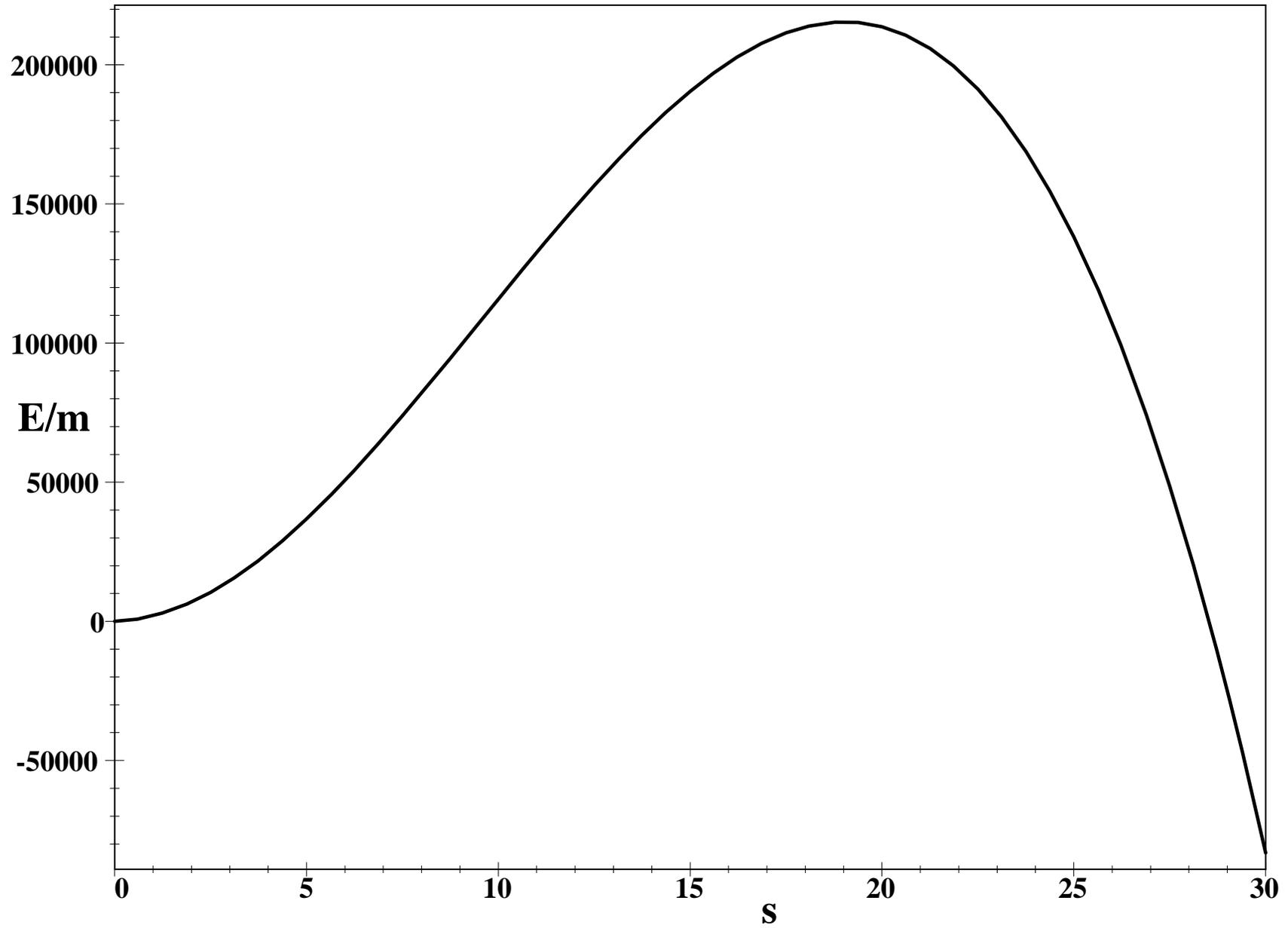
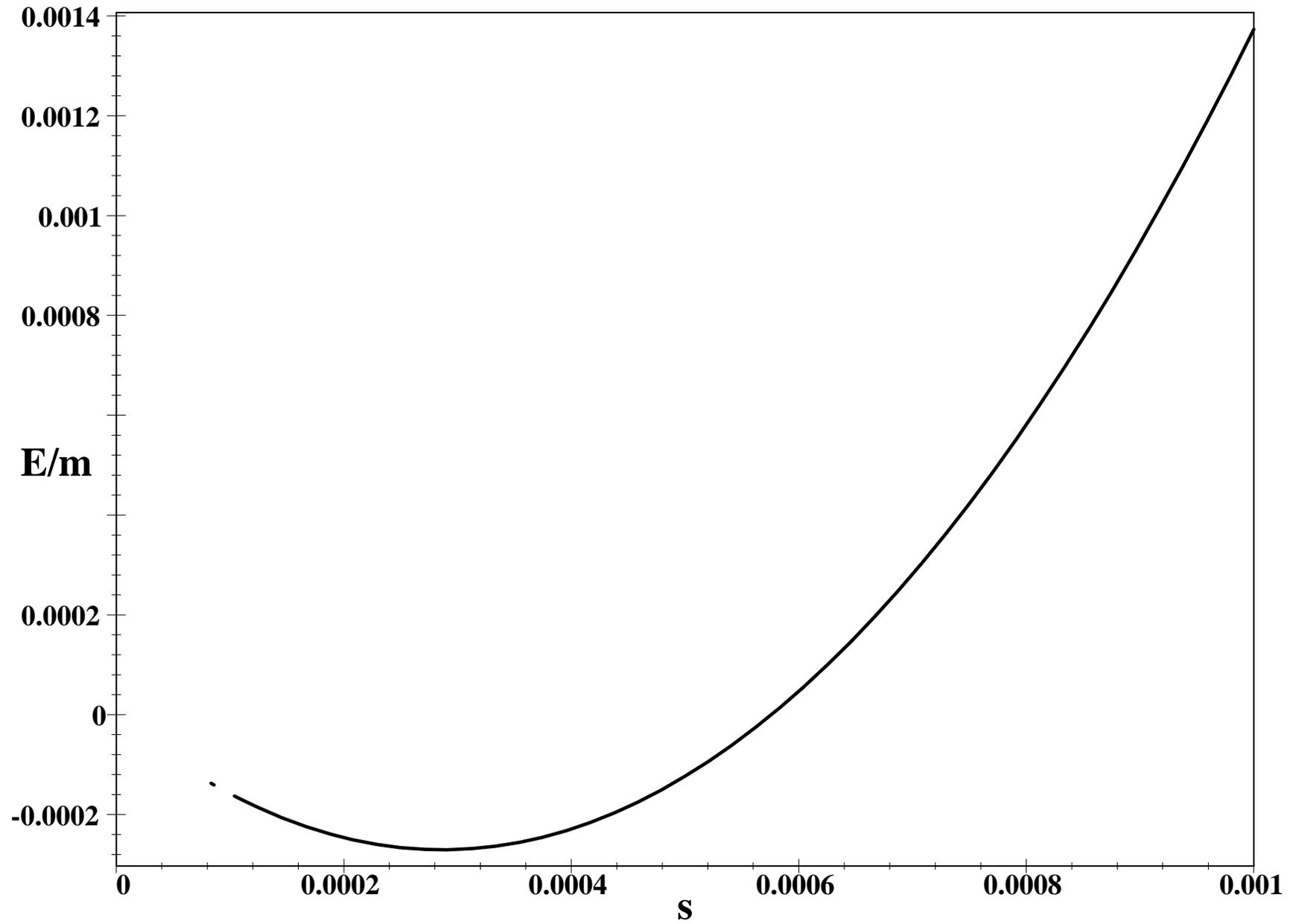
{\bf Figure 6}: Energy of the bubble as a function of the bubble radius
for $\nu=4$. (a) purely bosonic. (b) zoom of the interesting 
region for the approximate solution (including bound states).


\begin{thebibliography}{99}
\bibitem{nota} Here we used the trick of functionally differentiating 
and integrating
the effective action, i.e., we used the relation $\frac{\delta}{\delta M}tr[ln(i\gamma^{\mu}\partial_{\mu}+M(x))]={\cal G}$, where
${\cal G}$ is the Green function approximated by the gradient expansion.
\end{thebibliography}
\end{document}